\documentclass{article}
\usepackage{spconf,amsmath,graphicx,hyperref}
\usepackage{url}
\usepackage{wrapfig}
\usepackage[margin=1in]{geometry}
\usepackage{subcaption} 
\usepackage{float}      
\usepackage{amsthm}

\usepackage{booktabs}   
\usepackage{multirow}

\usepackage{amsmath,amsfonts,bm}

\def\eqref#1{equation~\ref{#1}}

\def\1{\bm{1}}

\def\vtheta{{\bm{\theta}}}

\def\vx{{\bm{x}}}
\def\vy{{\bm{y}}}

\DeclareMathAlphabet{\mathsfit}{\encodingdefault}{\sfdefault}{m}{sl}
\SetMathAlphabet{\mathsfit}{bold}{\encodingdefault}{\sfdefault}{bx}{n}

\title{SPADE: Structured Pruning and Adaptive Distillation \\ for Efficient LLM-TTS}

\name{Tan Dat Nguyen$^1$, Jaehun Kim$^1$, Ji-Hoon Kim$^1$, Shukjae Choi$^2$, Youshin Lim$^2$, Joon Son Chung$^1$}
\address{$^1$Korea Advanced Institute of Science and Technology, South Korea, $^2$42dot Inc., South Korea}

\begin{document}
%
\maketitle

\begin{abstract}
The goal of this paper is to introduce \textbf{SPADE}, a framework for \textbf{S}tructured \textbf{P}runing and \textbf{A}daptive \textbf{D}istillation for \textbf{E}fficient Large Language Model-based text-to-speech (LLM-TTS). Recent LLM-TTS systems achieve strong controllability and zero-shot generalization, but their large parameter counts and high latency limit real-world deployment. SPADE addresses this by combining (i) a \textit{pruning} step guided by a word-error-rate-based layer importance index to remove non-essential Transformer layers, with (ii) multi-level \textit{knowledge distillation} to restore autoregressive coherence. On zero-shot benchmarks, SPADE preserves near-parity perceptual quality while halving Transformer depth, reducing VRAM usage by up to $20\%$, and achieving up to $1.7\times$ faster real-time factor with less than $5\%$ of the original training data. These results show that compact LLM-TTS models can maintain naturalness and speaker similarity while enabling practical real-time speech generation. Audio samples are available at \footnote{\url{https://mm.kaist.ac.kr/projects/SPADE/}}.
\end{abstract}


\section{Introduction}
\label{sec:intro}

Large Language Model (LLM)-based text-to-speech (LLM-TTS) systems such as CosyVoice~\cite{du2024cosyvoice,du2025cosyvoice}, VALL-E~\cite{wang2023neural}, CLaM-TTS~\cite{kim2024clam}, RALL-E~\cite{xin2024rall}, and LLaSA~\cite{ye2025llasa} have shown advanced controllability, prosody modeling, and zero-shot generalization across speakers and languages. Early approaches trained LLM backbones directly on speech tokens~\cite{du2024cosyvoice, borsos2023audiolm}, while recent methods initialize from pretrained text LLMs (e.g., LLaMA~\cite{touvron2023llama}, Qwen2.5~\cite{bai2025qwen2}) and adapt them with speech objectives~\cite{ye2025llasa, zhang2023speechgpt}. Leveraging rich contextual representations, these systems synthesize natural speech conditioned on long prompts, speaker embeddings, and control tokens, pushing TTS closer to human-level performance.

Despite these advances, LLM-TTS models inherit the costly nature of text-only LLMs, including large parameter counts, high memory usage, and slow autoregressive decoding, and these factors are particularly pronounced in real-time deployment and on-device applications.
On the other hand, a line of research in text-only LLM domain has extensively studied model compression, including pruning~\cite{xu2021rethinking,liu2021ebert,gromov2024unreasonable}, distillation~\cite{sun2019patient,pan2020meta,ko2024distillm}, quantization~\cite{zafrir2019q8bert,yao2022zeroquant,huang2024slim}, and adaptive inference methods such as early exiting and token reduction~\cite{xin2020deebert,ma2023llm}. However, systematic compression methodology for LLM-TTS, where the preservation of prosody, naturalness, and long-context coherence serve as additional key aspects, remains underexplored. 

\begin{figure}[t]
    \centering
    \includegraphics[width=\linewidth]{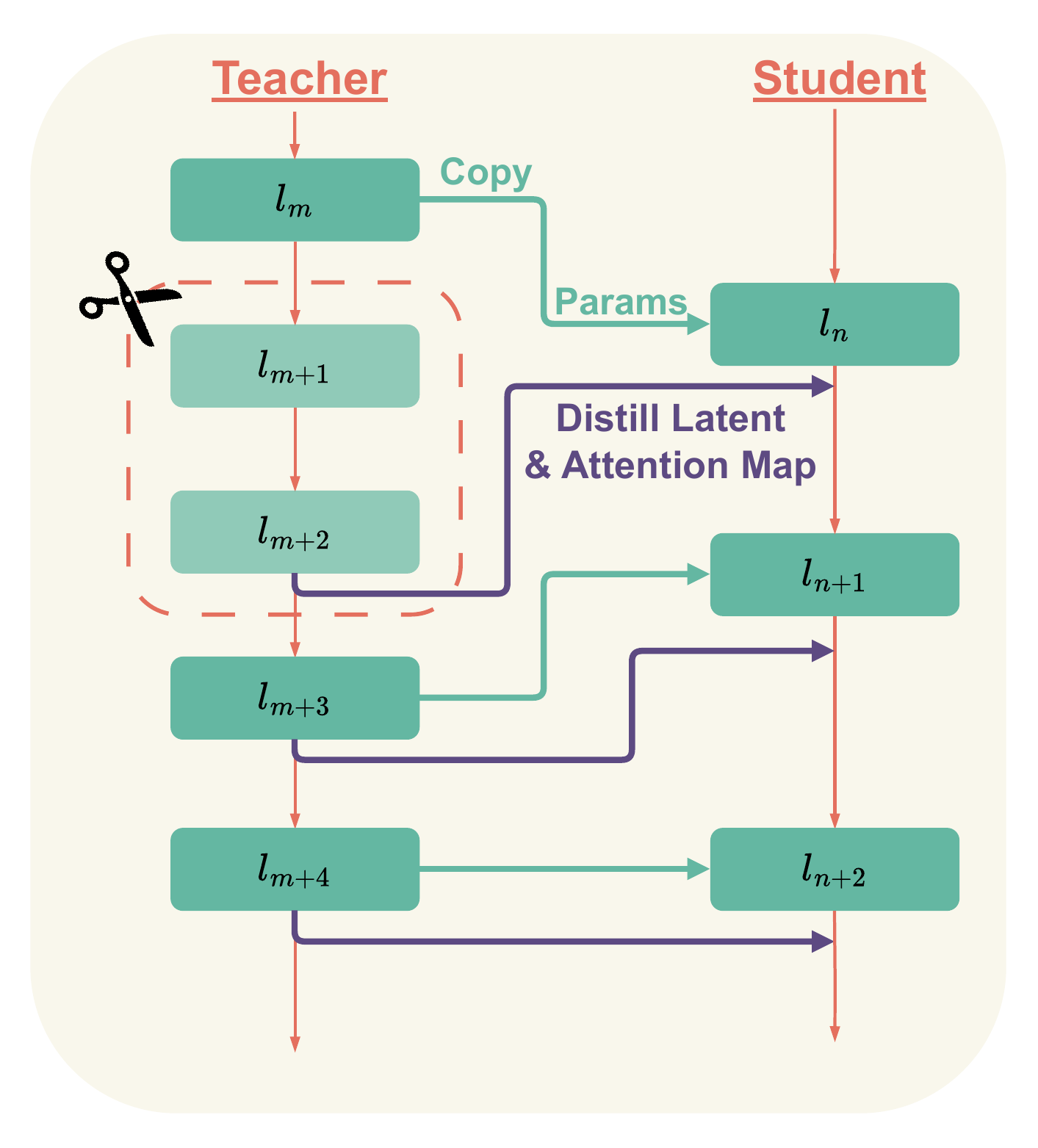}
    \caption{Overview of SPADE. A large LLM-TTS model is compressed into a smaller student model through \textit{pruning} and multi-level \textit{distillation}. Parameters are copied from retained layers, while latent states are aligned across pruned segments to preserve synthesis quality.}
    \vspace{-0.4cm}
    \label{fig:prun_and_heal}
\end{figure}

\begin{figure*}[t]
    \centering
    \begin{subfigure}[t]{0.49\textwidth}
        \centering
        \includegraphics[width=\linewidth]{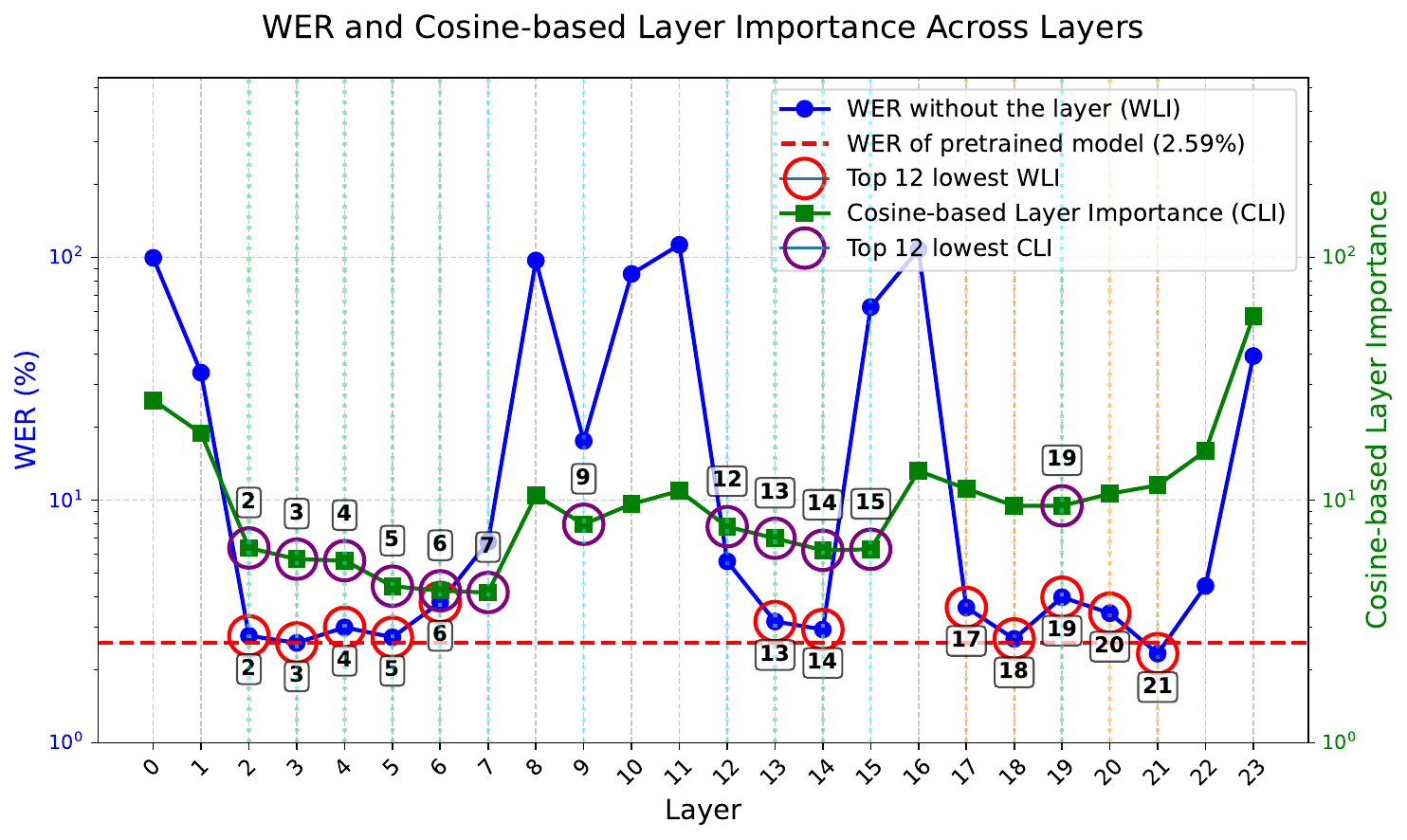}
        \caption{CosyVoice 2}
        \label{fig:transformer_graph}
    \end{subfigure}
    \hfill
    \begin{subfigure}[t]{0.49\textwidth}
        \centering
        \includegraphics[width=\linewidth]{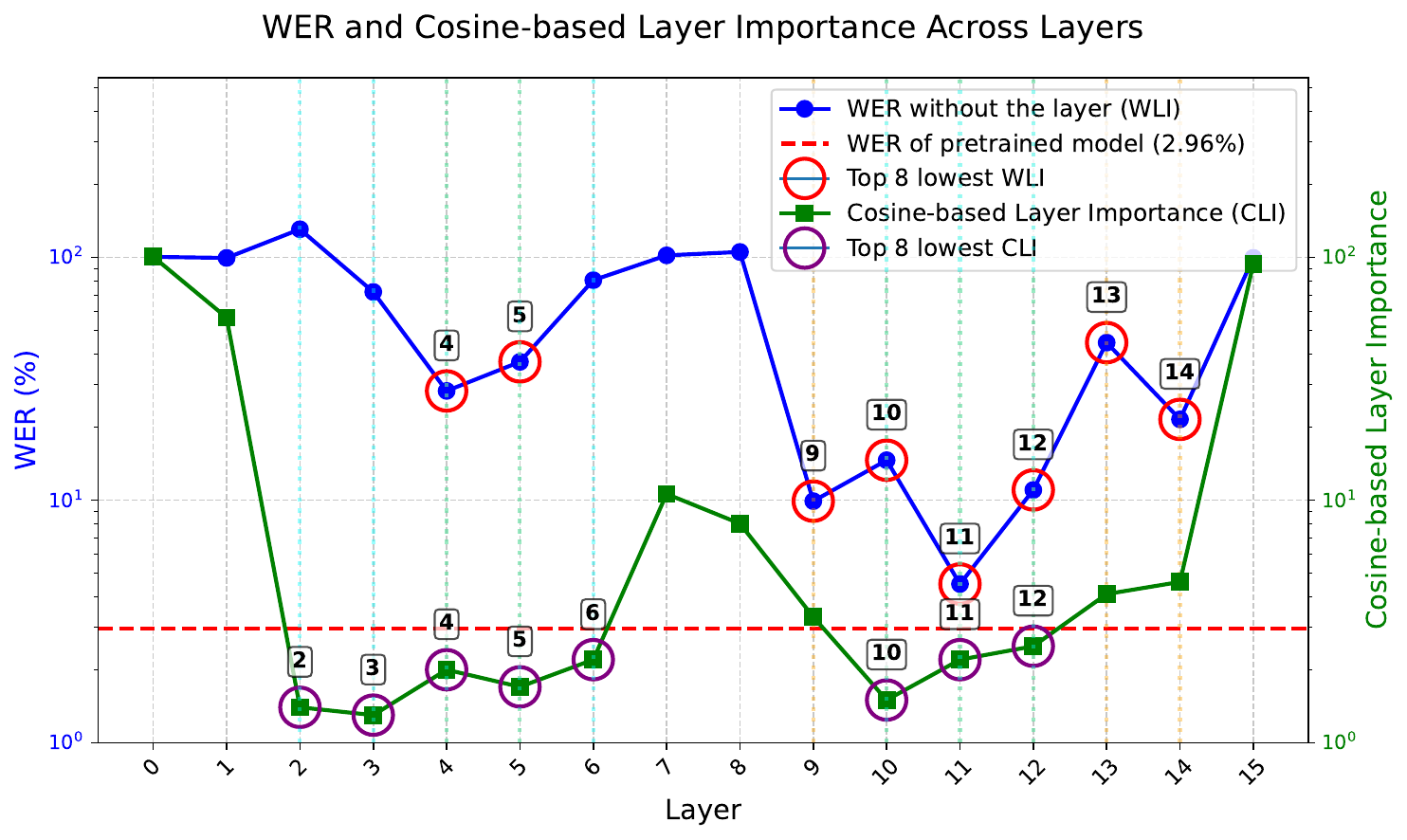}
        \caption{LLaSA}
        \label{fig:mamba_graph}
    \end{subfigure}
    \caption{
    WLI and cosine-based layer importance of (a) CosyVoice 2 and (b) LLaSA.
    High WLI indicates that WER increases significantly when the layer is removed, and high cosine-based importance indicates the input and output latents of the layer are dissimilar.
    We found that, based on WLI, the layers in the beginning, middle, and the end contribute critically to performance.
    Our method prunes the model by removing layers with least contribution to the performance.
    }
    \label{fig:wer_bi}
    \vspace{-0.4cm}
\end{figure*}

In this paper, we present \textbf{SPADE}, a framework for \textbf{S}tructured \textbf{P}runing and \textbf{A}daptive \textbf{D}istillation for \textbf{E}fficient LLM-TTS. By removing non-essential layers through a Word Error Rate (WER) based layer importance index and recovering performance via multi-level distillation, SPADE achieves substantial efficiency gains while preserving perceptual quality. Across zero-shot benchmarks, SPADE halves Transformer depth, reduces overall parameters by up to $40\%$, lowers VRAM consumption by up to $20\%$, and accelerates inference by as much as $1.7\times$, all while retaining near-parity in perceptual metrics. Moreover, the knowledge distillation process shows remarkably high data-efficiency: The recovery of performance requires less than $5\%$ of the pretraining data of the original checkpoints. These results highlight the framework of pruning and distillation as a practical pathway toward compact, high-fidelity, real-time speech generation.

\section{Proposed Method}
\label{sec:proposed_method}

Our framework first explores the importance of each layer in LLM backbone and prunes non-essential layers to compress the model.
Subsequently, an efficient knowledge distillation is applied to effectively restore the performance of the pruned model.

\subsection{Model compression through pruning}

SPADE is motivated by using the nature of residual connections: $x^{l} = x^{l-1} + f_l(x^{l-1})$, where $x^l$ and $f^l$ denote the hidden state and transformation in the layer $l$, respectively.
In LLM-TTS, each transformer layer contributes by refining latent representations through residual connections~\cite{he2016deep}.
Although the architecture shows strong performance in diverse applications, recent studies in text-based LLM suggest some layers provide only weak refinements and can be removed with little effect~\cite{ gromov2024unreasonable,veit2016residual}.
An established criterion to identify them is to compute the cosine distance between inputs and outputs of a layer~\cite{muralidharan2024compact}.
However, Fig.~\ref{fig:wer_bi} shows that cosine-based layer importance (CLI) does not align with the performance contribution of layers in TTS, as indicated by the different patterns in WER change.

To address this, we leverage WER, as the primary interest is the semantic consistency of generation, and propose WER-based layer importance (WLI):
\begin{equation}
    \mathrm{WLI}_{i} = \mathbb{E}_{\mathcal{D}} \Big[ \\
    \mathrm{WER}\big(\mathrm{model}(\vx_2; \vtheta_{\setminus i}, \vx_1, \vy_1), \vy_2\big)\Big],
\end{equation}
where ${\vtheta}_{\setminus i}$ denotes the model parameters without the $i$-th layer, $\mathcal{D}$ is a subset of the evaluation set, and $(\vx_1, \vy_1)$ and $(\vx_2, \vy_2)$ denote reference and query text-audio pairs, respectively.
Specifically, a layer is considered important only if its absence causes significant degradation in WER.
Unlike cosine-based layer importance, which only estimates the difference between inputs and outputs within a layer, WLI directly measures the contribution of each layer to the final performance.
As shown in Fig.~\ref{fig:wer_bi} evaluated WLI with Whisper~\cite{radford2022whisper}, and found that many of the layers have negligible impact on performance, indicating a significant redundancy and aligning with the findings in text-domain LLMs~\cite{gromov2024unreasonable}.
Moreover, we find that the earliest, central, and final layers consistently emerge as important across different LLM-TTS models.
Based on the analysis, we prune transformer layers with low WLI values from the LLM backbones.

\subsection{Recovering original performance}

While the proposed framework effectively reduces the parameters without additional modules, such as parameter-efficient fine-tuning~\cite{hu2022lora,dettmers2023qlora}, the pruned model naturally confronts disconnected flow of latent information.
To address this, we leverage the original un-pruned model as the teacher and perform a knowledge distillation training that simply \textit{heals} the pruned model to minimize the loss of performance without any additional parameters.
To maximize the restoration, we employ a composite loss that benefits from both supervised learning and teacher-guided knowledge distillation:
\begin{equation}
\mathcal{L} = \alpha * \mathcal{L}_{CE}
+ \frac{1 - \alpha}{4} *\big(
\mathcal{L}_{logit}
+ \mathcal{L}_{l}
+ \mathcal{L}_{a}
+ \mathcal{L}_{e}
\big).
\end{equation}

Here, the Cross-Entropy loss $\mathcal{L}_{CE}$ is responsible for the \textbf{supervised} component that directly guides the output distribution.
The \textbf{distillation} component comprises 4 elements: Embedding reconstruction loss $\mathcal{L}_{e}$, alignment losses on logit $\mathcal{L}_{logit}$, latent $\mathcal{L}_{l}$, and attention $\mathcal{L}_{a}$, following~\cite{muralidharan2024compact}.
To provide more stability, we implement the $\mathcal{L}_{logit}$ with mixed distribution by leveraging Skew KL Divergence~\cite{ko2024distillm}.
For $\mathcal{L}_{e}, \mathcal{L}_l$, and $\mathcal{L}_a$, we calculate Mean Squared Error (MSE) of embedding outputs, intermediate latents, and attention matrices between the teacher and student, respectively.
To maximize the distillation of teacher's knowledge, we propose to dynamically select layers to apply loss for $\mathcal{L}_l$ and $\mathcal{L}_a$.
Specifically, as shown in Fig.~\ref{fig:prun_and_heal}, the target values (latent, attention map) for the student layer ($l_n$) are derived from the last layer before the next retained layer ($l_{m+2}$) from the teacher model.
While simple, this approach allows the pruned student model to retain not only its original capability but also those of the removed layers, culminating in a smaller yet more compact model.
Finally, we combine the supervised and distillation components by adjustable weight $alpha$ to balance the influence of each, where the value is empirically set to $0.25$ to deliberately provide stronger supervised guidance.

\section{Experimental Setup}
\label{sec:results}

To evaluate whether pruning and distillation can truly enable compact yet high-fidelity LLM-TTS systems, we benchmark SPADE against two representative baselines: CosyVoice 2\footnote{\url{https://github.com/FunAudioLLM/CosyVoice}}~\cite{du2024cosyvoice2} and LLaSA-1B\footnote{\url{https://github.com/zhenye234/LLaSA_training}}~\cite{ye2025llasa}. Both models are chosen for their strong zero-shot capabilities and public availability of checkpoints.

Since a key motivation of SPADE is to minimize retraining cost, we fine-tune each pruned variant on only a fraction of the data: $25\%$ of LibriHeavy~\cite{kang2024libriheavy} (EN) for LLaSA and LibriTTS~\cite{zen2019libritts} (EN) for CosyVoice 2. This corresponds to less than $5\%$ of the pretraining corpus size, testing whether our framework can recover quality under data-constrained conditions.
 
We assess the performance on \textit{LibriTTS test-clean} and the \textit{Seed-TTS eval set}~\cite{anastassiou2024seed}, both of which are widely used for zero-shot evaluation. All experiments are conducted on 4$\times$NVIDIA A6000 GPUs, with official training scripts left unchanged aside from data size and pruning. CosyVoice 2 is fine-tuned with dynamic batches up to 20,000 tokens, whereas LLaSA uses a batch size of 4. Fine-tuning with SPADE runs for 7 epochs to CosyVoice 2 and for 1 epoch when applying it to LLaSA.

To evaluate both computational efficiency and perceptual quality, we consider a range of complementary metrics. Efficiency is assessed in terms of model depth, parameter count, and real-time factor (RTF), while intelligibility is measured using word error rate (WER). Perceptual aspects are captured objectively through speaker similarity (SS) and UTMOS using VERSA toolkit~\cite{shi2024versaversatileevaluationtoolkit}, and subjectively through the naturalness mean opinion score (NMOS) with 20 listening and 50 random samples from both evaluation sets per model. Together, these metrics reveal whether compact models preserve the qualities essential for real-world TTS deployment.

\begin{table*}[t]
\centering
\small

\begin{subtable}{\textwidth}
\centering
\resizebox{1.0\textwidth}{!}{\begin{tabular}{lcccccccccc}
\toprule
\multirow{2}{*}{\textbf{Model}} & 
\multirow{2}{*}{\textbf{Layers}} & 
\multirow{2}{*}{\textbf{Params}} & 
\multirow{2}{*}{\textbf{RTF}} & 
\multirow{2}{*}{\textbf{NMOS}} &
\multicolumn{3}{c}{\textbf{Seed-TTS \textit{test-en}}} &
\multicolumn{3}{c}{\textbf{LibriTTS \textit{test-clean}}} \\
\cmidrule(lr){6-8} \cmidrule(lr){9-11}
& & & & & \textbf{WER} & \textbf{SS} & \textbf{UTMOS}&
\textbf{WER}& \textbf{SS} & \textbf{UTMOS} \\
\midrule
Human Record    & -  & -    & -     & $3.96 \pm 0.14$ & $1.47$   & $1.00$   & $3.52$   & $1.85$   & $1.00$ & $4.14$ \\
\midrule
Vocoder Resyn. of GT  & -  & -    & -     & $3.87 \pm 0.15$ & $1.53$   & $1.00$   & $3.53$   & $1.65$   & $1.00$   & $4.01$   \\
CosyVoice 2~\footnotemark[4]~\cite{du2024cosyvoice2}       & $24$ & $0.63$  & $0.61$ & $3.71 \pm 0.13$ & ${2.03}$   & ${0.66}$   & $4.15$   & ${1.43}$  & $0.81$ & ${4.41}$ \\
CosyVoice 2 + Ours & ${12}$ & ${0.38}$ & ${0.35}$ & ${3.58\pm 0.14}$ & $2.71$   & ${0.66}$   & ${4.16}$   & $1.59$  & ${0.82}$   & ${4.41}$ \\
CosyVoice 2 + Ours & ${9}$ & ${0.32}$ & ${0.33}$ & ${3.55\pm 0.14}$ & $3.09$   & ${0.66}$   & ${4.15}$   & $1.94$  & ${0.81}$ & ${4.40}$ \\
\midrule
Codec Resyn. of GT  & -  & -    & -     & $3.59\pm0.15$ & $2.49$   & $1.00$   & $3.69$   & $2.52$   & $1.00$   & $4.01$   \\
LLaSA~\footnotemark[4]~\cite{ye2025llasa}           & $16$ & $1.7$  & $0.82$ & ${3.37\pm0.15}$ & ${3.54}$   & ${ 0.46}$   & ${ 4.13}$   & ${1.54}$   & ${ 0.47}$ & $4.41$ \\
LLaSA + Ours     & ${8}$  & ${ 1.3}$ & ${ 0.58}$ & $3.11\pm0.14$ & $4.20$   & $0.41$   & $4.06$   & $1.88$   & $0.43$ & ${ 4.40}$ \\
\bottomrule

\end{tabular}}

\caption{Absolute zero-shot evaluation results on Seed-TTS Eval set and LibriTTS test-clean. 
Higher values indicate better performance for NMOS, SS, and UTMOS, while lower values indicate better performance for RTF and WER. 
Missing entries correspond to trivial cases, e.g., human records or codec/vocoder resynthesis of groundtruth (GT).
}
\label{tab:abs_results}
\end{subtable}

\vspace{0.8em}

\begin{subtable}{\textwidth}
\centering
\resizebox{1.0\textwidth}{!}{\begin{tabular}{lccccccc}
\toprule
\textbf{Model Pair} & 
\textbf{Layers} & 
\textbf{Params} & 
\textbf{RTF} & 
\textbf{Data} & 
\textbf{WER} & 
\textbf{SS} & 
\textbf{UTMOS} \\

\midrule
CosyVoice2 $\rightarrow$ CosyVoice2 + Ours (12)
& $\downarrow 50.0\%$ 
& $\downarrow 39.7\%$ 
& $\downarrow 42.6\%$ 
& $2(0.3)\%$ $\downarrow$ 
& $+0.43$ $\uparrow$ 
& $+0.005$ $\uparrow$ 
& $0.00$ $\leftrightarrow$ \\
CosyVoice2 $\rightarrow$ CosyVoice2 + Ours (9)
& $\downarrow 62.5\%$ 
& $\downarrow 49.2\%$ 
& $\downarrow 45.9\%$ 
& $2(0.3)\%$ $\downarrow$ 
& $+0.79$ $\uparrow$ 
& $+0.000$ $\leftrightarrow$ 
& $0.00$ $\leftrightarrow$ \\
\midrule
LLaSA $\rightarrow$ LLaSA + Ours 
& $\downarrow 50.0\%$ 
& $\downarrow 23.5\%$ 
& $\downarrow 29.3\%$ 
& $\leq12.5(5)\%$ $\downarrow$ 
& $+0.50$ $\uparrow$ 
& $-0.045$ $\downarrow$ 
& $-0.04$ $\downarrow$ \\
\bottomrule
\end{tabular}}
\caption{Relative performance of SPADE models compared to their uncompressed versions. 
For data usage, the exact amount of English data in the internal LLaSA pretraining set is unknown; we therefore report an upper bound ($\leq$).}
\label{tab:gap_results}
\end{subtable}

\caption{Zero-shot evaluation on Seed-TTS (\textit{test-en}) and LibriTTS (\textit{test-clean}). Table~\ref{tab:abs_results} shows efficiency and quality, while Table~\ref{tab:gap_results} reports relative gaps of pruned variants.}
\label{tab:combined_results}
\vspace{-0.5cm}
\end{table*}

\begin{figure}
    \centering
    \includegraphics[width=\linewidth]{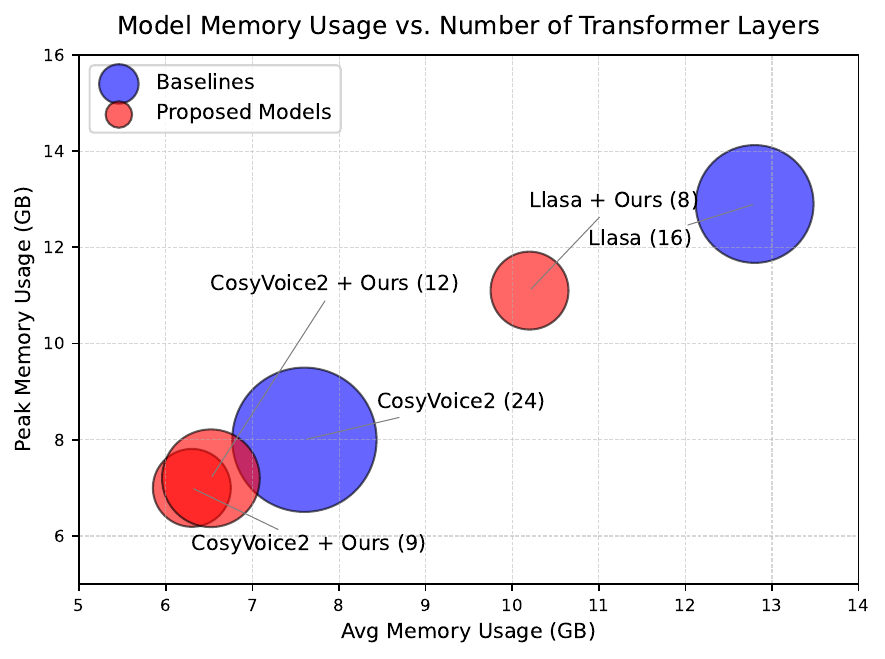}
    \caption{SPADE effectively reduces half of the Transformer layers, reducing VRAM usage by $14\%$ for CosyVoice 2 and $20\%$ for LLaSA.}
    \label{fig:model_size}
    \vspace{-0.4cm}
\end{figure}

\section{Results and Analysis}

\subsection{Effect on performance after pruning}
Table~\ref{tab:combined_results} presents the main results of applying SPADE to CosyVoice 2 and LLaSA with different configurations.
For CosyVoice, pruning to 12 layers halves the depth, reduces parameters by $39.7\%$ and accelerates inference by $42.6\%$.
Moreover, effective VRAM usage is reduced by $14\%$, as depicted in Fig.~\ref{fig:model_size}.
In particular, these gains come with no significant degradation in both quantitative and qualitative metrics, showing a slight increase of 0.68 in WER for the challenging Seed-TTS dataset and a 0.11 reduction in NMOS.
A more aggressive variant with only 9 layers further reduces parameters by $49.2\%$, improves RTF by $45.9\%$, and lowers VRAM usage by $17\%$.
While this extreme setting renders additional increase in WER, perceptual metrics such as NMOS, SS, and UTMOS remain stable, suggesting that SPADE enables flexible trade-offs between efficiency and intelligibility.
To confirm that our framework generalizes beyond a single backbone, we conduct experiment on LLaSA, a larger variant based on speech codec.
Here, pruning removes half the layers, decreases parameters by $23.5\%$, improves RTF by $29.3\%$ ($1.41\times$ speed-up), and reduces VRAM by $20\%$.
The result suggests that, while most of the metrics lie in acceptable range, the performance degradation is relatively larger compared to CosyVoice 2.
Based on the analysis in Fig.~\ref{fig:wer_bi}\footnote{Metrics are reported based on official checkpoints}, we expect it attributes to overall high WLI values across all layers, indicating each layer contributes similarly to the performance.

\subsection{Ablation study}

We evaluate the effectiveness of the proposed methodology with a systematic ablation study.
Table~\ref{tab:ablation} presents experiments on CosyVoice 2 with LibriTTS test-clean.
First, the proposed WLI-based pruning is replaced with existing cosine-based pruning.
The result shows a notable increase in WER and CER, demonstrating the pruning based on WLI, a metric directly related to intelligibility, successfully prevents performance degradation.
Moreover, dynamic distillation loss is removed and student layer is distilled only with the information from the corresponding teacher layer.
The overall decrease in performance suggests that adaptively choosing the target in distillation training is simple yet plays a significant role.

\begin{table}[t]
\centering
\resizebox{0.48\textwidth}{!}{\begin{tabular}{lcccc}
\toprule
\textbf{Expr.} & 
\textbf{WER $\downarrow$} & 
\textbf{CER $\downarrow$} & 
\textbf{SS $\uparrow$} & 
\textbf{UTMOS $\uparrow$} \\

\midrule
CosyVoice 2
& $1.43$ 
& $0.46$ 
& $0.81$ 
& $4.41$ \\

\midrule
CosyVoice 2 + Ours
& $1.59$ 
& $0.54$ 
& $0.82$ 
& $4.41$  \\

Cosine-based pruning
& $1.74$
& $0.61$
& $0.81$
& $4.40$ \\

Distill from original layer
& $1.65$ 
& $0.58$
& $0.81$
& $4.40$ \\

\bottomrule
\end{tabular}}
\caption{
Ablation experiment on LibriTTS test-clean.
Both cosine-based pruning and the alternative latent knowledge distillation scheme degrade the overall performance, where cosine-based pruning shows more significant increase in WER and CER.
}
\label{tab:ablation}
\vspace{-0.4cm}
\end{table}

\section{Conclusion}

We presented SPADE, a pruning-based framework for compressing LLM-TTS models, and showed that SPADE achieves substantial efficiency gains while preserving intelligibility and naturalness. We investigate the importance of each layer using the proposed WLI and find that many layers contribute little to audio synthesis.
Experimental results verify that, by applying SPADE, such layers can be removed without harming perceived quality.

\small{\section{acknowledgement}
This work was partially supported by Institute of Information \& communications Technology Planning \&
Evaluation (IITP) grant funded by the Korean government (MSIT, RS-2025-02263977, Development of Communication Platform supporting User Anonymization and Finger Spelling-Based Input
Interface for Protecting the Privacy of Deaf Individuals).}

\small

\bibliographystyle{IEEEbib}
\bibliography{strings,shortstrings,refs}

\end{document}